\documentclass[12pt]{article}
\usepackage{pdproc}
\usepackage{graphicx}

\begin{document}

\renewcommand{\thefootnote}{\alph{footnote}}
  
\title{
 STATUS OF NEUTRINO FACTORY R\&D WITHIN THE MUON COLLABORATION}

\author{ RAJENDRAN RAJA}

\address{ Fermilab\\
 P.O.Box 500, Batavia, Illinois 60510, U.S.A\\
 {\rm E-mail: raja@fnal.gov}}

\abstract{
We describe the current status of the research within the Muon
Collaboration towards realizing a Neutrino Factory. We describe
briefly the physics motivation behind the neutrino factory approach to
studying neutrino oscillations and the longer term goal of building
the Muon Collider. The benefits of a step by step staged approach of
building a proton driver, collecting and cooling muons followed by the
acceleration and storage of cooled muons are emphasized. Several usages
of cooled muons open up at each new stage in such an approach and new 
physics opportunites are realized at the completion of each  stage.}
   
\normalsize\baselineskip=15pt

\section{Introduction}
\begin{figure}[tbh]
\centerline{\includegraphics[width=4.0in,angle=-90]{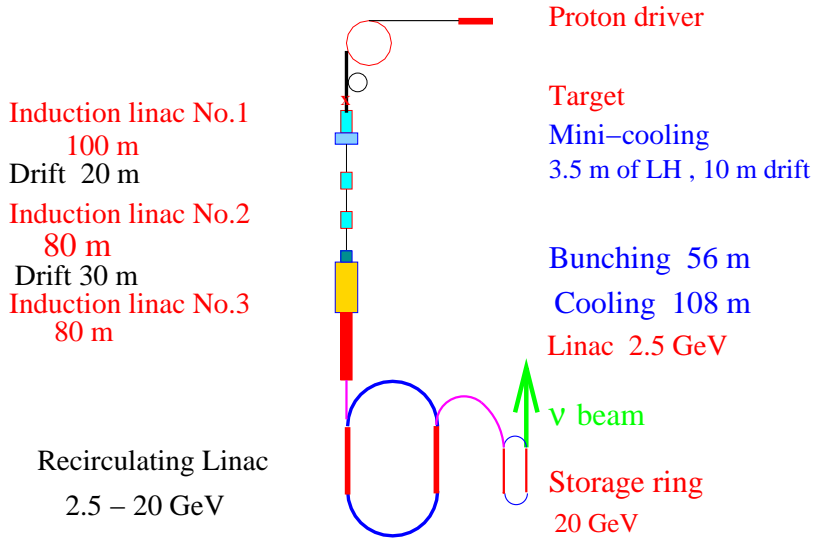}}
\caption[Schematic of the Neutrino Factory Study-II version]{Schematic of
the Neutrino Factory Study-II version.}
\label{nufact-scheme-bnl}
\end{figure}

 The Neutrino Factory and Muon Collider Collaboration, also known as
 the Muon Collaboration is an international organization consisting of
 $\approx$ 130 physicists from 36 institutions whose mission is to
 design and build the Neutrino Factory. The collaboration was
 initially established to study ways to make the Muon Collider a
 reality, by collecting, cooling and accelerating intense beams of
 muons of both charges. When evidence mounted  for the existence of
 neutrino oscillations in the late 1990's, the potential of a ring
 storing a high intensity beam of muons in producing a focused beam of
 neutrinos (and anti-neutrinos) in studying neutrino oscillations was
 realized~\cite{king,rajageer,geer} 
 and the collaboration shifted its efforts towards the study
 of Neutrino Factories. In order to produce the muons, one needs an
 intense proton source (known as the proton driver), a scheme to
 collect the pions produced by bombarding a mercury target with the
 protons, a scheme to cool the muons so that they can be accelerated
 and a scheme to accelerate the muons rapidly and store them in a
 storage ring with long straight sections where the muons decay to
 produce the neutrino beams that are directed at the neutrino
 detectors placed at a suitable distance from the
 source. Figure~\ref{nufact-scheme-bnl} is the schematic for such a
 machine.

 In this report, we outline the physics potential of a Neutrino
 Factory and the Muon Collider and describe the R\&D efforts being
 made within the Muon Collaboration towards realizing these
 goals. Detailed discussion of the topics covered here may be found in
 the status reports published by the 
 collaboration~\cite{statrep1,statrep2,study1,study2}.

\section{The Physics Potential of the Neutrino Factory and the Muon Collider}
The stored muons in a muon storage ring decay into electrons and
neutrinos (e.g $\mu^+\rightarrow e^+\nu_e\bar\nu_\mu$). The energy
spectrum of the neutrinos is exactly calculable using the Standard
Model. Since the electron mass is much smaller than the muon mass, the
neutrinos carry off, on average, the same amount of energy as the
electron. This is in stark contrast to the beta beams case~\cite{beta}, 
where the
heavy ion decay products carry off most of the energy of the initial
state heavy ion and the neutrinos are at much lower energy than the
beam energy.  

\begin{figure}[tbh!]
\centerline{\includegraphics[width=4.0in]{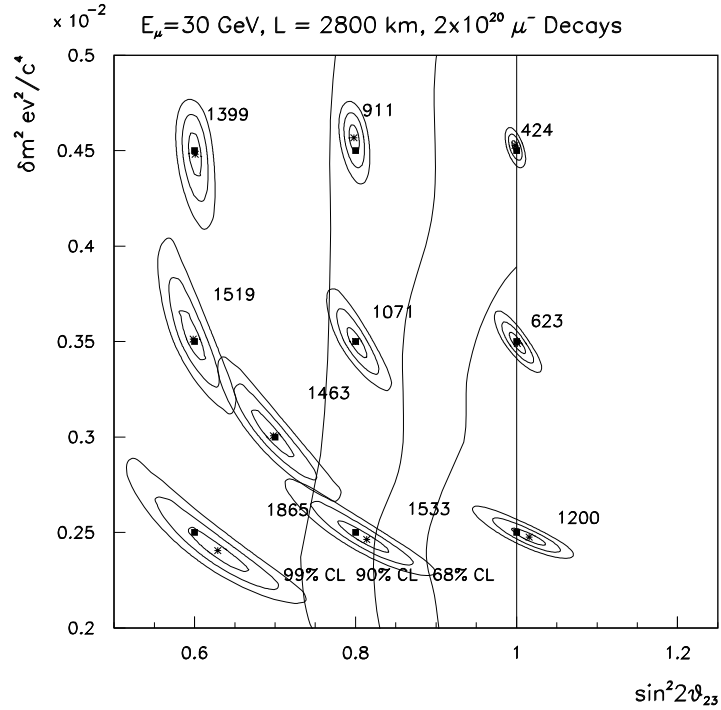}}
\bigskip
\caption[Error ellipses in  $\delta m^2$ sin$^2 2\theta$ space for a Neutrino Factory]
{ \label{fig:30gev_disap_fit}
Fit to muon neutrino survival distribution for $E_\mu=30$ GeV and $L=2800$~km for 10
pairs of sin$^2 2\theta$, $\delta m^2$ values. For each fit, the
1$\sigma$,\ 2$\sigma$
and 3$\sigma$ contours are shown. The generated points are indicated by the
dark
rectangles and the fitted values by stars. The SuperK 68\%, 90\%, and 99\% 
confidence
levels are superimposed. Each point is labelled by the predicted number of 
signal events for that point.}
\end{figure}

Figure~\ref{fig:30gev_disap_fit} shows the error
contours in the $\delta m^2_{32}$ sin$^2 2\theta_{23}$ space for a
30~GeV Neutrino Factory with $2\times 10^{20}~\mu^-$ decays and a 10
kT detector~\cite{barger-long}. 
The precision acheivable using a neutrino factory in
measuring these parameters far exceeds the presently available
precison using atmospheric neutrino detectors such as Super-K as well
as those likely to be available using superbeams in the near future.

\begin{figure}[tbh!]
\centerline{\includegraphics[width=4.0in]{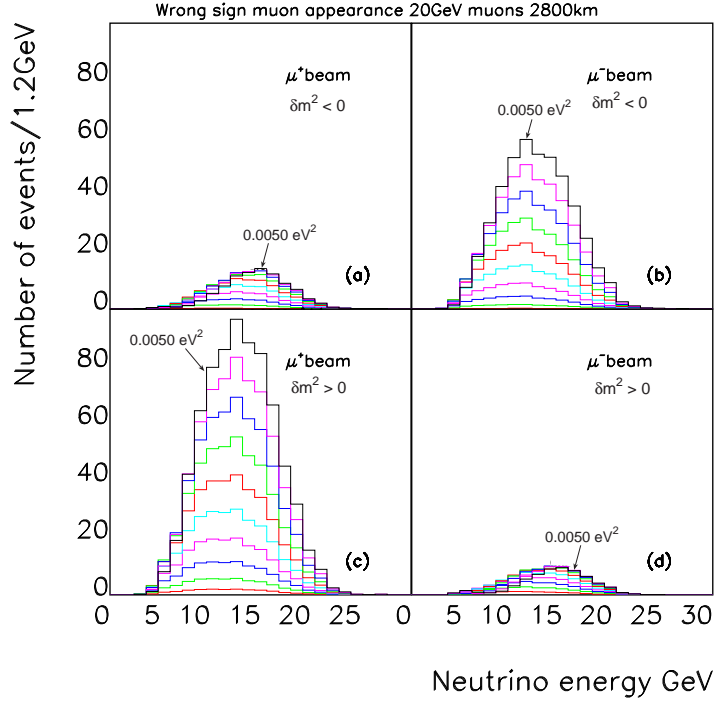}}
\caption[Wrong sign muon appearance rates and sign of  $\delta m^2_{32}$]
{The wrong sign muon appearance rates for a 20 GeV muon storage ring at
a baseline of 2800~km with 10$^{20}$ decays and a 50 kiloton detector
for (a)~$\mu^+$ stored and negative $\delta m^2_{32}$\,, (b)~$\mu^-$ stored
and negative $\delta m^2_{32}$\,, (c)~$\mu^+$ stored and positive $\delta
m^2_{32}$\,,
(d)~$\mu^-$ stored and positive $\delta m^2_{32}$. The values of $|\delta
m^2_{32}|$ range from 0.0005 to 0.0050 eV$^2$ in steps of 0.0005~eV$^2$.  
Matter enhancements are evident in (b) and (c).
\label{fig:hists}}
\end{figure}

\begin{figure}[tbh!]
\centerline{\includegraphics[width=4.0in]{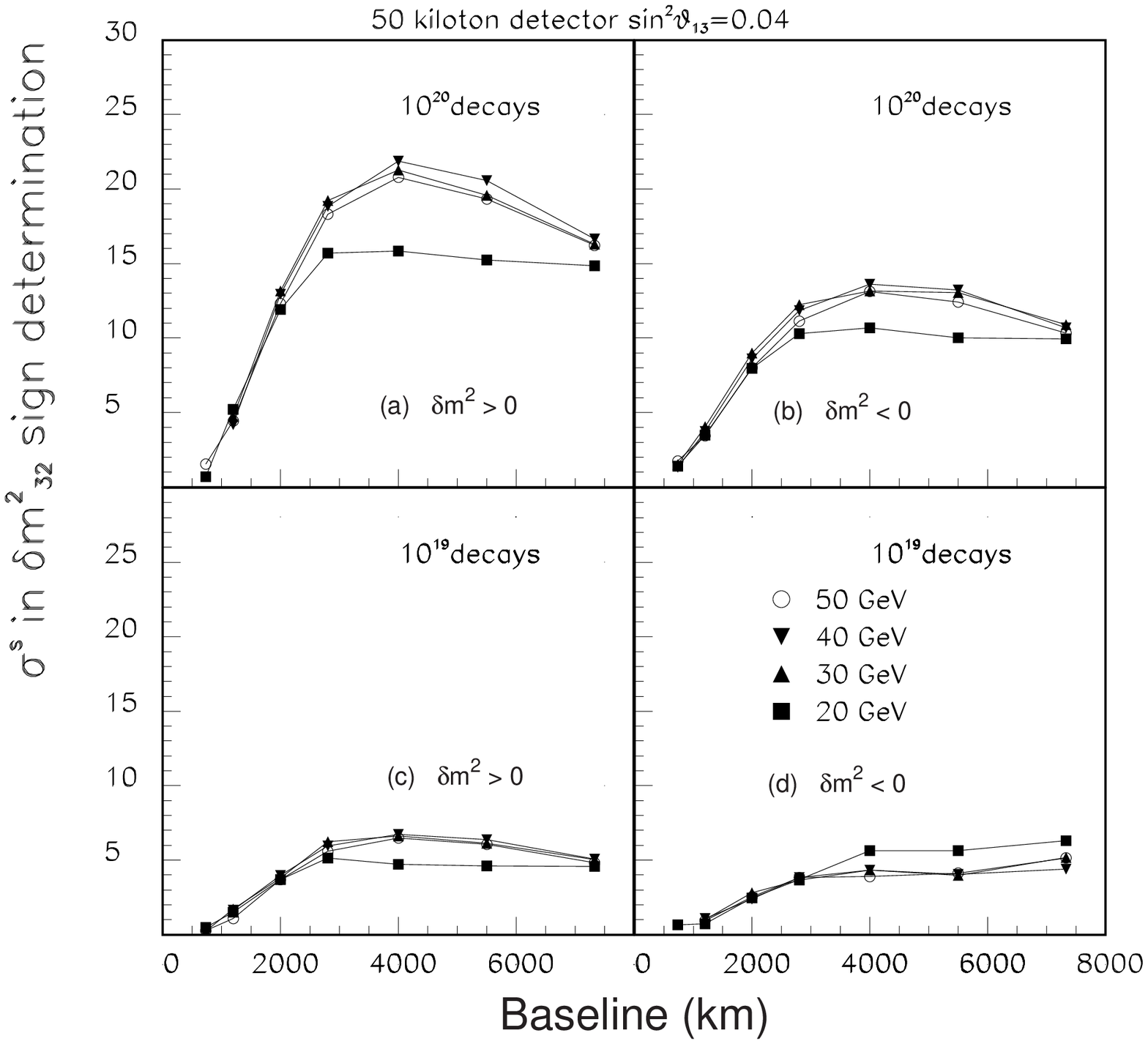}}
\caption[$\delta m_{32}^2$ sign determination at a Neutrino Factory]
{The statistical significance (number of standard deviations) 
with which the
sign of $\delta m_{32}^2$ can be determined versus baseline length for
various muon storage ring energies. The results are shown for 
a 50~kiloton detector, and (a)~10$^{20}$
$\mu^+$ and $\mu^-$ decays and positive values of $\delta m_{32}^2$;
(b)~10$^{20}$ $\mu^+$ and $\mu^-$ decays and
negative values of $\delta m_{32}^2$; (c)~10$^{19}$ $\mu^+$ and 
$\mu^-$ decays and positive values of $\delta
m_{32}^2$; (d)~10$^{19}$ $\mu^+$ and $\mu^-$
decays and negative values of $\delta m_{32}^2$.
\label{fig:sigmas}}
\end{figure}

Neutrino factories permit long baseline experiments ($>$ 2000~km) and
this enables matter effects to be investigated with some precision.
Figure~\ref{fig:hists} shows the energy spectra of wrong sign muons
that appear~\cite{barger-sign} 
due to the oscillation of electron to muon neutrinos in a
detector with a baseline of 2800 km for a 20 GeV Neutrino Factory with
both positive and negative muons. If $\delta m^2_{32}>0$, one gets 
an enhancement for the oscillation ($\nu_e\rightarrow \nu_\mu$)~ $\rightarrow
\mu^-$ due to matter effects (Figure~\ref{fig:hists}(c)) and the
reverse is the case for the case $\delta m^2_{32}<0$, as shown
in Figure~\ref{fig:hists}(b). The results are for a 50 kiloton detector and 
$10^{20}$ muon decays. The value of sin$^2 2\theta_{13}$=0.04  is assumed 
for this simulation as well as the best LMA values available at the time 
of publication of the paper.
Figure~\ref{fig:sigmas} shows the number of standard deviations~\cite{barger-sign} with
which one can distinguish the sign of $\delta m^2_{32}$ for an entry
level Neutrino Factory with $10^{19}$ muon decays as well as a
Neutrino Factory with $10^{20}$ decays as a function of baseline. 
Even for an entry level
Neutrino Factory, it would be possible to distinguish 
the sign of $\delta m^2_{32}$ at the $3\sigma$ level for baselines 
of the order of 2800 km.

\begin{figure}[tbh!]
\centerline{\includegraphics[width=4.0in]{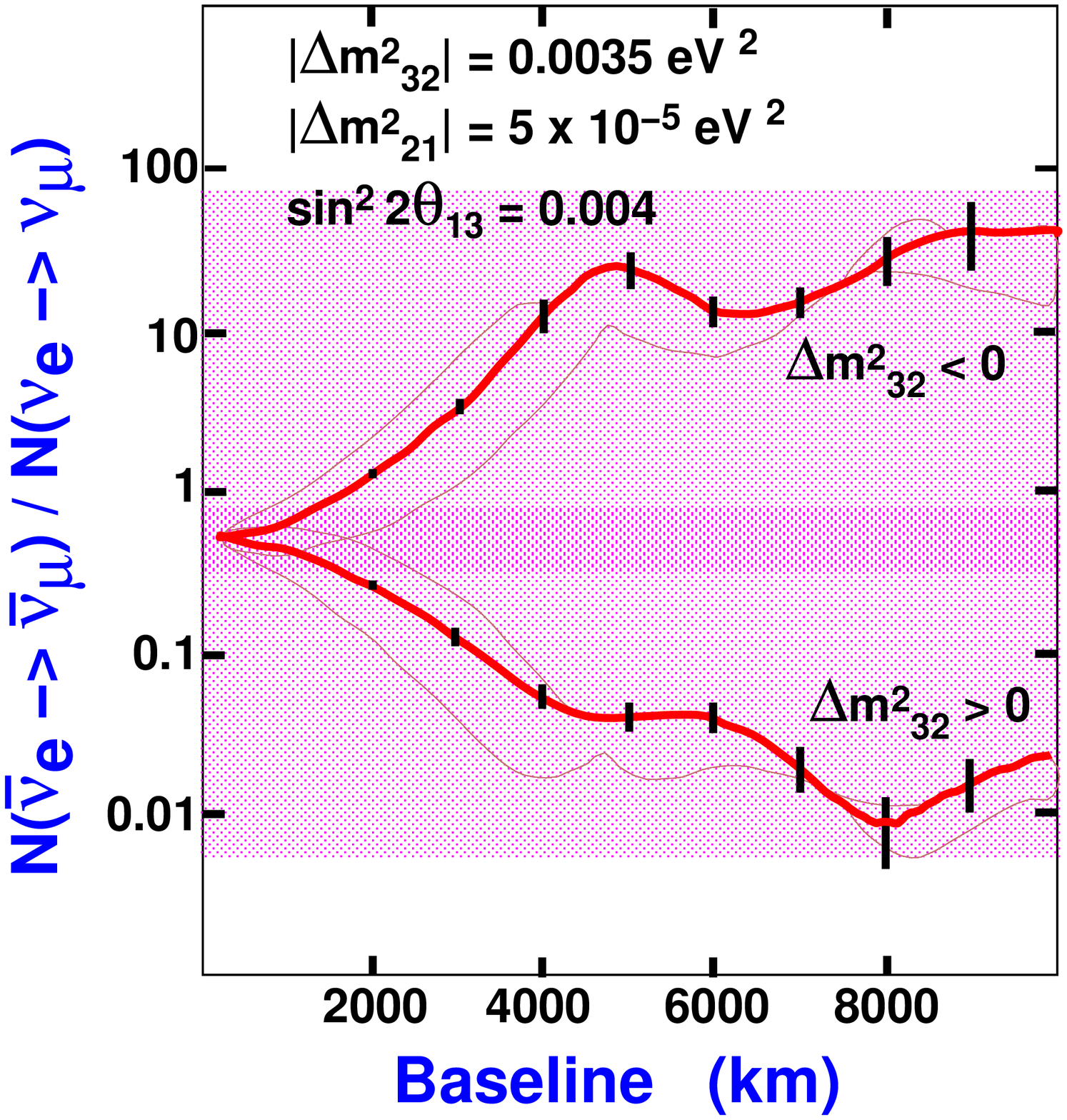}}
\bigskip
\caption[CP violation effects in a Neutrino Factory]
{ \label{cpfig}
Predicted ratios of wrong-sign muon event rates when positive and
negative muons are stored in a 20~GeV Neutrino Factory, shown as a
function of baseline.  A muon measurement threshold of 4~GeV is
assumed. The lower and upper bands correspond, respectively, to negatve
and positive $\delta m^2_{32}$. The widths of the bands show how the
predictions vary as the $CP$ violating phase $\delta$ is varied from
$-\pi$/2 to $\pi$/2, with the thick lines showing the predictions for
$\delta=0$. The statistical error bars correspond to a
high-performance Neutrino Factory yielding a data sample of 10$^{21}$
decays with a 50~kiloton detector. The curves are based on calculations
presented in~\cite{barger-entry}. }
\end{figure}

Having determined the sign of $\delta m^2_{32}$, the CP
violation parameter $\delta$ can be determined. Figure~\ref{cpfig}
shows the predicted ratios of wrong-sign muon event rates when 20~GeV 
positive and negative muons are stored in a Neutrino Factory as a function 
of baseline. The shaded region indicates the change in the rates as the CP 
violating phase $\delta$ is varied from $-\pi/2$ to $\pi/2$ and the error 
bars indicate the statistcal errors in the data. A value of 
sin$^2 2\theta_{13}$=0.004 is assumed for this plot. It can be seen that the 
Neutrino Factory is capable of exploring the structure of neutrino mixing by 
measuring currently unknown parameters ($sin^2 2\theta_{13}$ and $\delta$) 
over ranges unmatched by other techniques and the known parameters 
($\sin^2 2\theta_{23}, \delta m^2_{32}$) with much better precision 
than other methods.

 It is also possible to do precision neutrino non-oscillation physics
at the Neutrino Factory~\cite{mangano}. Partons densities with $x>0.1$
can be measured with an order of magnitude more precision than
currently available. Polarized parton densities become measurable and
sin$^2\theta_{W}$ can be measured with an error of $\approx 2\times
10^{-4}$. Hydrogen targets can be employed, bypassing nuclear effects
and rare lepton flavor violating decays of muons can be tagged with
the appearance of wrong sign electrons and muons or prompt taus.

 The energy of a muon beam in a Neutrino Factory or a muon collider
can be determined to an accuracy of a part per million using $g-2$
precession~\cite{rajatol} and this can be put to good use in a Muon
Collider with a center of mass energy equal to the Higgs boson mass,
also known as a Higgs Factory. Higgs bosons can be produced in the $s$
channel in a Muon Collider, since the cross section is $\approx$
40,000 times than that present in an $e^+e^-$ collider, by virtue of
the mass of the muon being $\approx 200$ times that of the
electron. The accurate determination of the energy of a muon bunch in
a Muon Collider permits the scanning of the Higgs boson peak and a
determination of its width. In a Minimal Supersymmetric version of the
Standard model which has two sets of degenerate heavy Higgs bosons (H
and A), the Muon Collider can be used to tell them apart by means of
an $s$ channel scan. The Muon Collider is compact and fits on existing
laboratory sites. If the problems related to cooling and acceleration
can be solved, it represents a means to reach energies as high as 2-3
TeV in the center of mass.
\section{Proton Driver}
\begin{figure}[tbh]
\begin{center}
\includegraphics[width=5.5in]{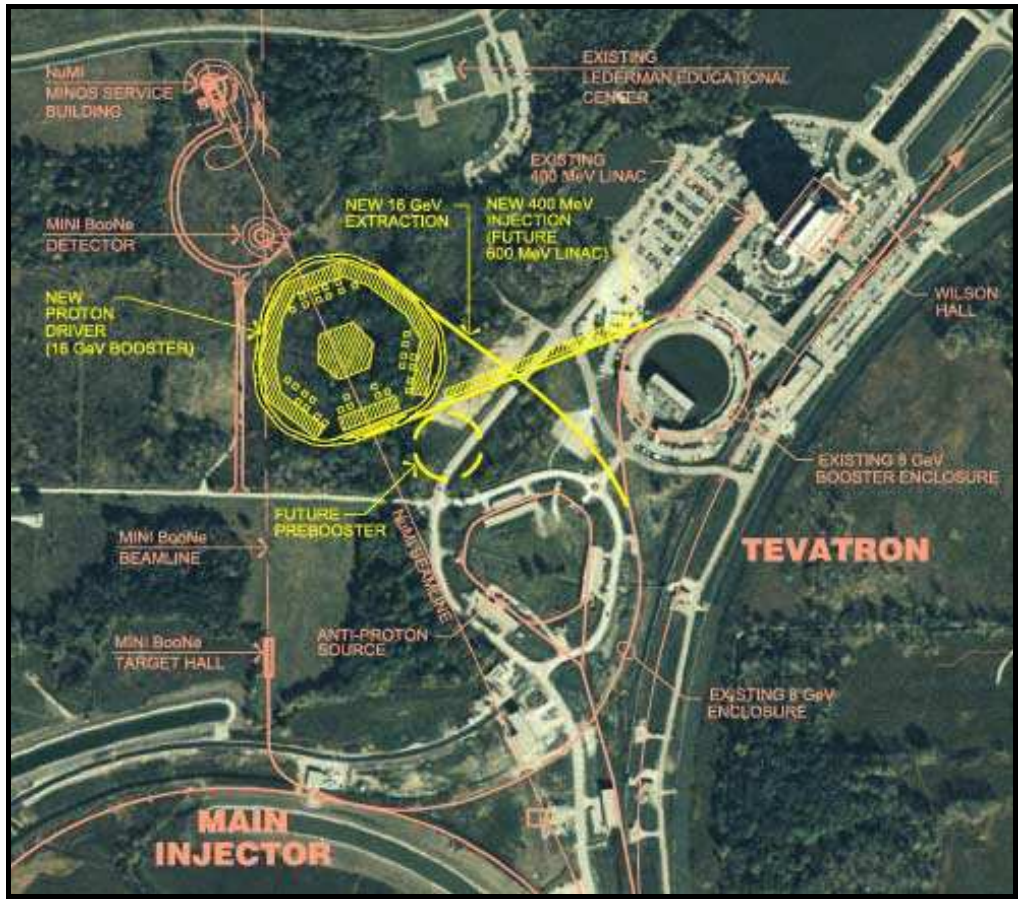}
\end{center}
\caption{FNAL proton driver layout from Ref. \protect\cite{FNALbooster}.}
\label{Proton:fnal}
\end{figure}

\begin{figure}[tbh]
\begin{center}
\includegraphics[width=5.5in]{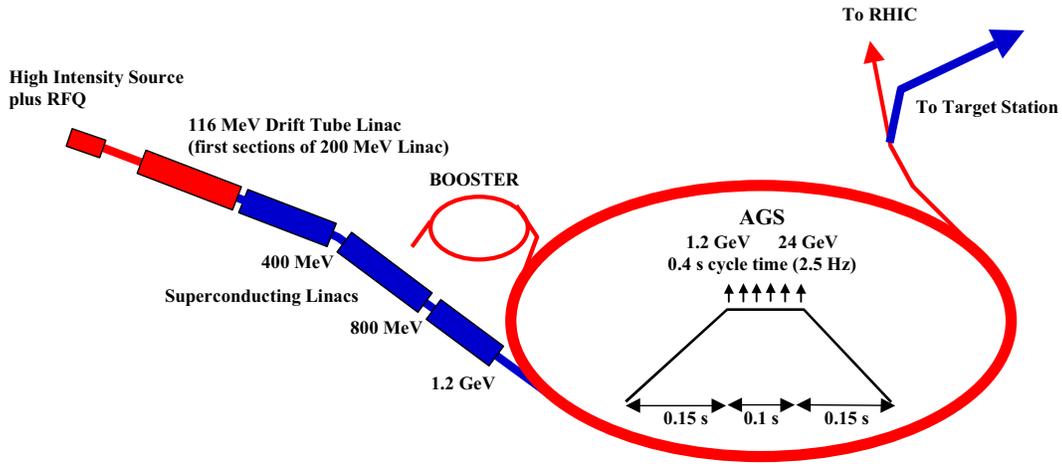}
\end{center}
\caption{AGS proton driver layout.}
\label{Proton:bnl}
\end{figure}

We now describe the stages involved in realizing a Neutrino Factory
and a Muon Collider. The first stage calls for an intense beam of
protons and necessitates the building of proton driver. The total
power of such a proton machine needs to be of the order of 4MW using current 
schemes of collecting and cooling muons. The building  of a 
1MW proton driver is expected to cost \$250M-330M and a 4MW driver is 
expected to cost \$330M-410M. Figure~\ref{Proton:fnal} shows the layout 
of such an object on the Fermilab site and Figure~\ref{Proton:bnl} shows 
a corresponding scheme on the Brookhaven site. 
Such a proton driver has multiple uses. It can be used to produce
conventional neutrino beams of high intensity (known in the jargon as
superbeams) to extend the measurement of the oscillation parameters
beyond current limits. Superbeam sensitivity to oscillation parameters
has been studied extensively compared to Neutrino Factories and is
inferior to that achievable with Neutrino
Factories~\cite{barger-super}. Various groups at
CERN~\cite{cern-proton}, Brookhaven~\cite{study2} and
Fermilab~\cite{FNALbooster} are currently exploring several different
designs for building such a machine.
\section{Producing and collecting muons}

\begin{figure}[tbh]
\begin{center}
\includegraphics*[width=4in]{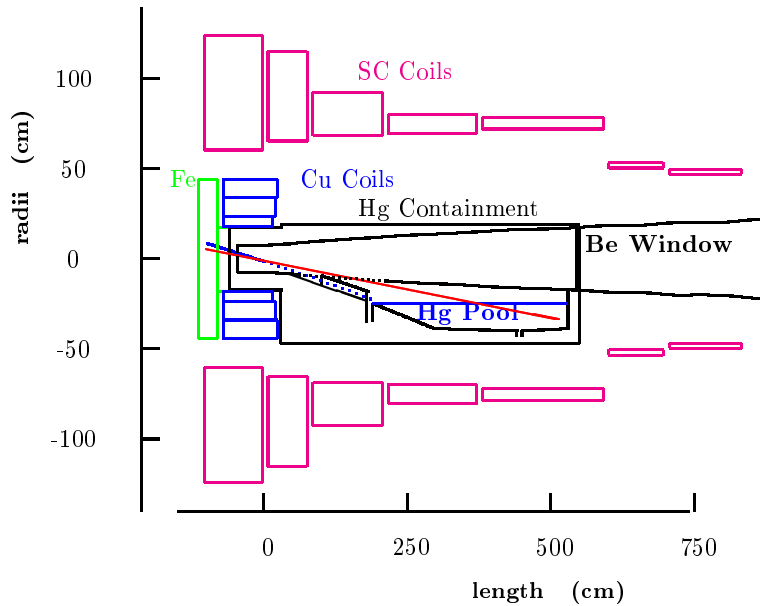}
\end{center}
\caption[Target, capture solenoids and mercury containment ]{Target, capture
solenoids and mercury containment.}
\label{tgtc}
\end{figure}

The intense proton beam from the proton driver is focused on a target
(currently liquid mercury, within the Muon Collaboration), which
produces large numbers of pions. The pions are contained in a 20T
solenoid inside which sits the target as shown in Figure~\ref{tgtc}. The high magnetic field is needed to confine  the interaction
products which are produced with large spreads in energy and angle. 
The pions from the interaction are at various transverse and longitudinal
momenta. They need to be ``phase rotated'' and bunched. The term
``phase rotation'' is used to describe the process by which a beam of
 pions of varying energy is acted on by electric fields that vary
with time and position along the line of flight of the pion to produce
a beam that is more uniform in energy.

\begin{figure}[tbh]
  \centering
  \includegraphics*[width=4in]{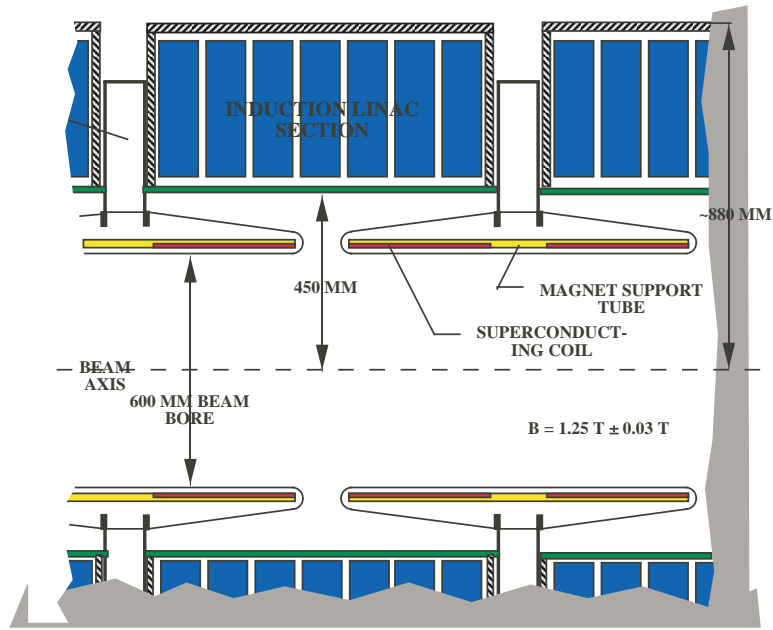}
  \caption[Induction cell and mini-cooling solenoid]{Cross section of the
    induction cell and transport solenoids.}
  \label{ind-lin}
\end{figure}

There are several schemes that have been investigated to do phase
rotation.  Figure~\ref{ind-lin} illustrates the induction linac scheme
used~\cite{study2} in study II.  After phase rotation, the beam goes into an rf
buncher which imposes an rf structure on the beam. The collection,
phase rotation and bunching channels collectively act as a decay
region where pions decay into muons. These muons need to be cooled so
that they can be accelerated and stored relatively easily.  Due to the
short lifetime of the muons ($\approx 2 \mu s$), the conventional
cooling techniques such as stochastic cooling or electron cooling are
inapplicable. Ionization cooling, whereby the muons lose energy in
low-Z absorbers (such as liquid hydrogen), and the lost longitudinal
energy is compensated by rf acceleration, at present provided the only
means to cool muons. The momentum of the muons at the end of the decay
channel is $\approx$ 250~MeV/c and the rf frequency is 200~MHz. The
transverse emittance of the beam at the end of the decay channel is
12mm-rad.  For the neutrino factory, the cooling requirements are
less stringent than for the muon collider, and it is only necessary to
cool this emittance to $\approx 2.7$~mm-rad.

\begin{figure}[tbh] 
\includegraphics[height=3.6in]{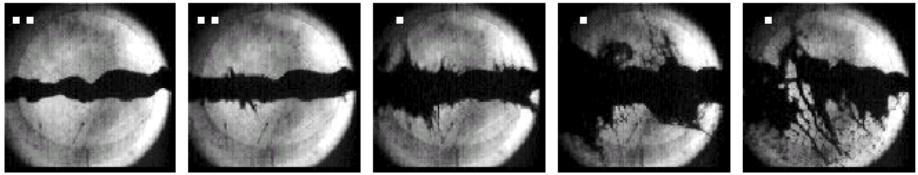}
\caption{Targetry setup and photographs of beam hitting mercury jets at
 t=0,0.75,2,7 and 18 ms. 
\label{targets}}
\end{figure}

Currently, the collaboration is engaged in building and testing a 
15 Tesla version of the high 
field solenoid housing the mercury target. Tests have also been done
with 1-cm diameter mercury jets exposed to beam~\cite{bnl-target} 
of 2E12 protons
as seen in Figure~\ref{targets}. The dispersal of the jet due to the interaction with the beam was not found to be destructive.
\section{Ionization Cooling}
\begin{figure}[hbt!]
\centering
\includegraphics[width=4in]{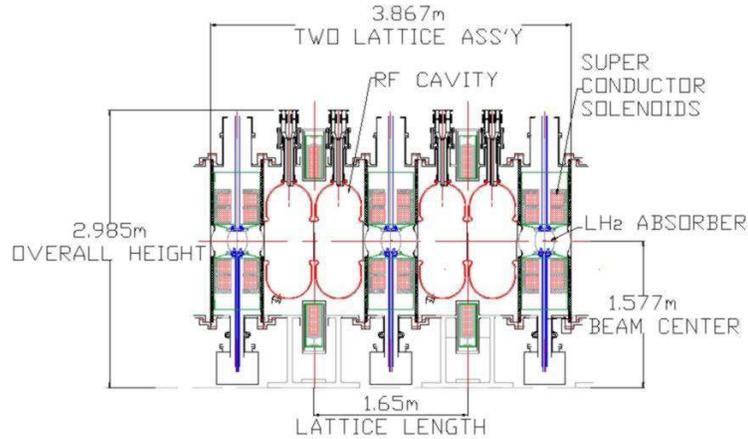}
\caption{Two cells of the 1.65 m cooling lattice.}
\label{coollat}
\end{figure}

Solenoidal cells provide large acceptance to the beam at the end of
the buncher.  They focus the beam at the absorber. It is important that
the divergence of the beam be as large as possible at the absorber so
as to minimize the emittance growth due to multiple scattering which
adds to the angular divergence of the beam in quadrature. The absorber
of choice currently is liquid hydrogen. This is to minimize the
effects due to multiple scattering for a given energy loss.  Each cell
of the cooling lattice contains three solenoids. Figure~\ref{coollat}
shows two cells of such a lattice and the placement of the liquid
hydrogen absorber and the rf modules.  The direction of the solenoids
reverses in alternating cells to prevent the buildup of canonical
angular momentum in the cooled beam. Each solenoid in this lattice
design (known as SFOFO in the jargon) is 3-5 Tesla in strength and the
rf frequency is 201MHz.  The energy loss per absorber is 7-12 MeV and
the rf voltage must be adjusted to compensate for this energy loss. 
Every cooling
channel is characterized by a number called the ``equilibrium
emittance'', which is that emittance at which the cooling rate equals
the rate at which the particles are heated by multiple scattering and
straggling.  The channel is ``tapered'' in that as the cooling
proceeds, one  increases the strength of cooling by making
the solenoidal focusing larger and the equilibrium emittance smaller.

\begin{figure}[tbh]
\centering
\includegraphics*[width=100mm]{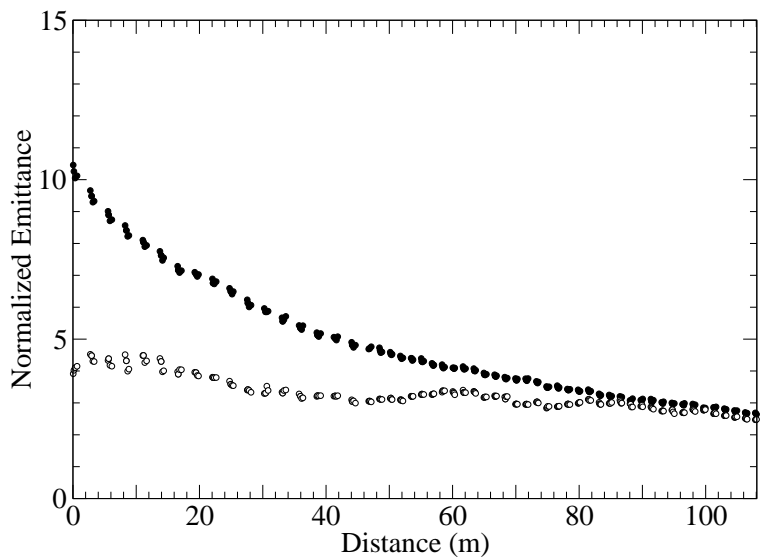}
\caption[The transverse and longitudinal emittances]{The transverse
  (filled circles, in mm) and longitudinal (open circles, in cm)
  emittances, as a function of the distance down the cooling channel.}
\label{EmittCool}
\end{figure}

 Figure~\ref{EmittCool} shows the transverse and longitudinal emittance
of the beam as a function of distance down this cooling channel. We
replace the lost energy in the longitudinal direction using the rf
modules and the beam cools transversely. However, these cooling
channels do not cool in the longitudinal direction, since energy
fluctuations due to straggling cause heating in the longitudinal
direction.

\begin{figure}[tbh]
\centering
\includegraphics*[width=100mm]{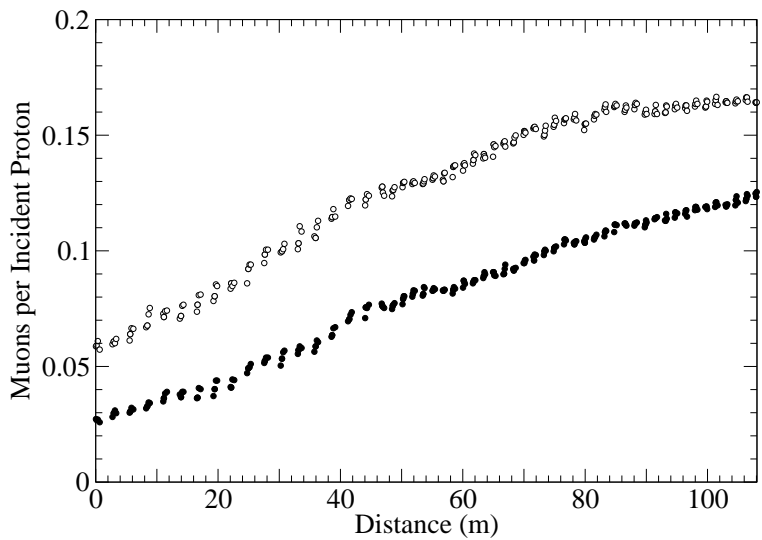}
\caption[$\protect\mu /p$ yield ratio for the two transverse emittance
cuts]{Muons per incident proton in the cooling channel that would
  fall within a normalized transverse acceptance of 15~mm (open circles) or
  9.75~mm (filled circles).}
\label{YieldCool}
\end{figure}

The figure of merit at the end of a cooling channel for a Neutrino
Factory is the number of muons/proton that will fit into the
acceptance of the accelerator. Figure~\ref{YieldCool} shows the number
of muons/proton that will be admitted into  accelerator acceptances
of 9.75~mm and 15~mm as a function of length of the cooling
section. These are rising functions of cooling section length
indicating the efficacy of the cooling process.

The price tag of the collection, phase rotation and cooling section is
estimated to be \$660M-840M. It is expected to produce a cold muon
source of $4\times 10^{20}$ muons per year with an energy spread of
~4.5\%. There are a large number of ``spin-off'' uses for such a
source of cold muons. One can use it to study rare decay modes of the
muon, and after a modest acceleration to the ``magic energy'' of 3.1 GeV, 
measure the $g-2$ of the muon to high precision as well as search for the
electric dipole moment of the muon. Cold muons can also be used for
muon radiography for medical as well as other applications. For a 
comprehensive review of the uses of cold muons see the 
status report~\cite{statrep2}.

\begin{figure}[tbh] 
\includegraphics[height=3.6in]{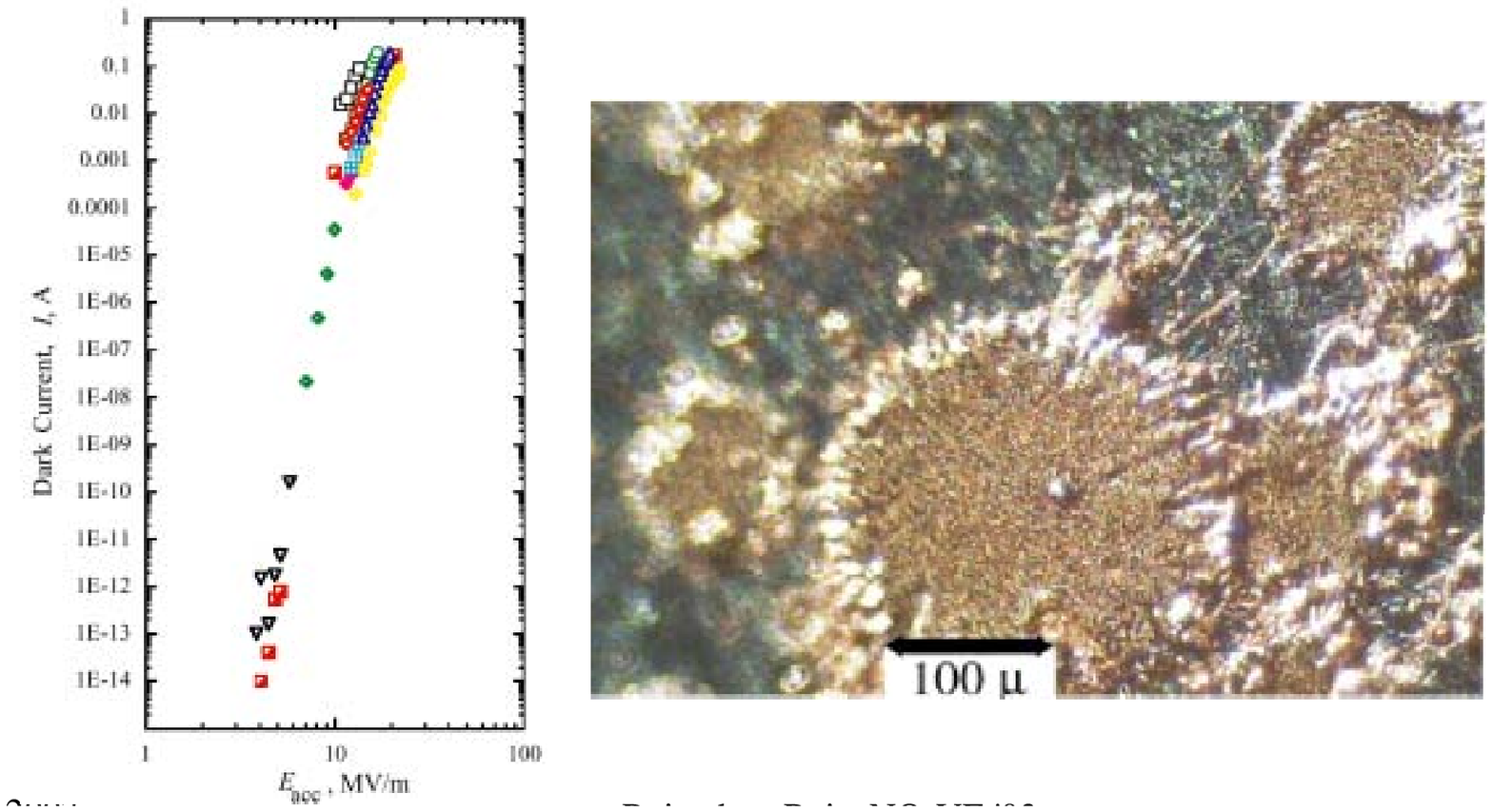}
\caption{Dark current measurements in rf cavity in a magnetic field. Damage to 
the cavity can be seen.
\label{dark}}
\end{figure}

The collaboration is engaged in R\&D efforts in fabricating liquid
hydrogen absorbers~\cite{absb}, fabricating and testing warm and
superconducting rf cavities, and in mounting an international effort
to demonstrate ionization cooling of muons, called
MICE~\cite{blondel}.  The warm rf work has been conducted for 805 MHz
cavities and also for 201MHz cavities. When the warm rf is put in a
magnetic field, damage to the cavity has been observed due to
breakdown currents. Figure~\ref{dark} shows the dark current
measurements as a function of rf voltage as well as photomicrographs
of the damage to the cavity.  More work is being done to understand
and control these currents. One of the avenues being explored is the
use of high pressure gas such as hydrogen to contain these
discharges. An rf workshop~\cite{rf-norem} was held at Argonne
National Laboratory during the fall of 2003 to address these issues.

\begin{figure}[tbh] 
\includegraphics[width=5.5in]{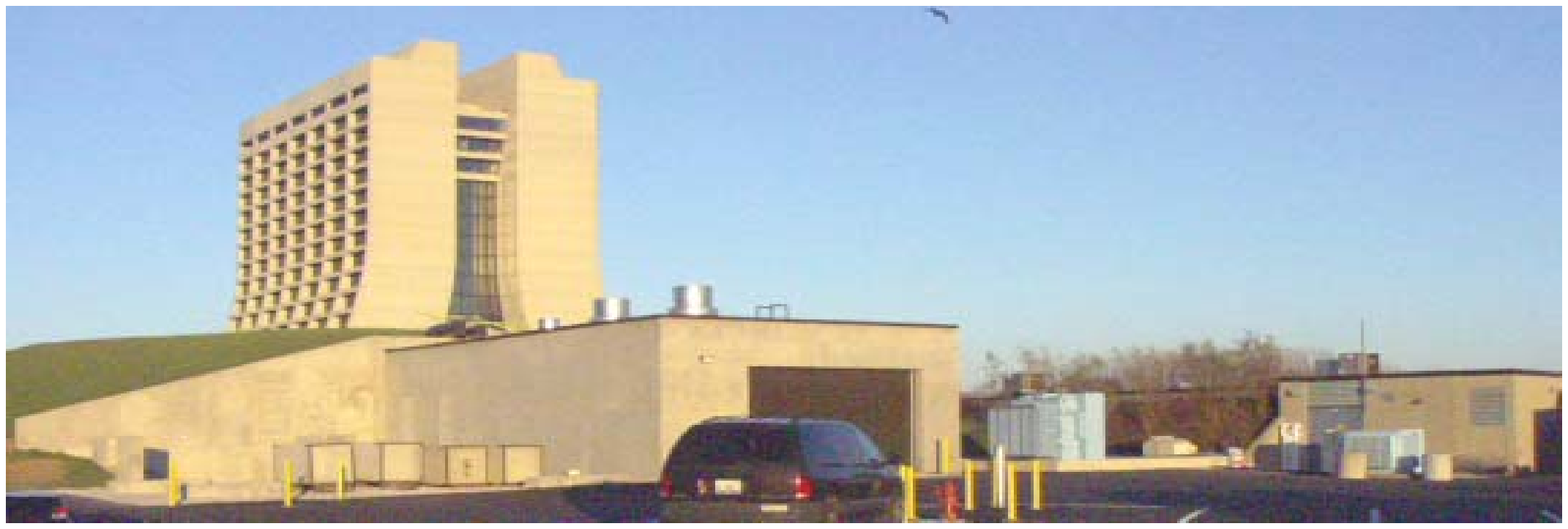}
\caption{The newly completed Mucool Test Area at the end of the Fermilab Linac.
\label{mta}}
\end{figure}

In addition, the collaboration has managed to construct the Mucool
Test Area at the end of the Fermilab Linac (See Figure~\ref{mta})
which will be used to test absorbers and rf modules by exposing them
to high intensity linac beams (400 MeV $H^-$).

\section{Acceleration and storage ring}
\begin{figure}[tbh]
\vspace{15mm}
  \centering
  \includegraphics[width=\textwidth]{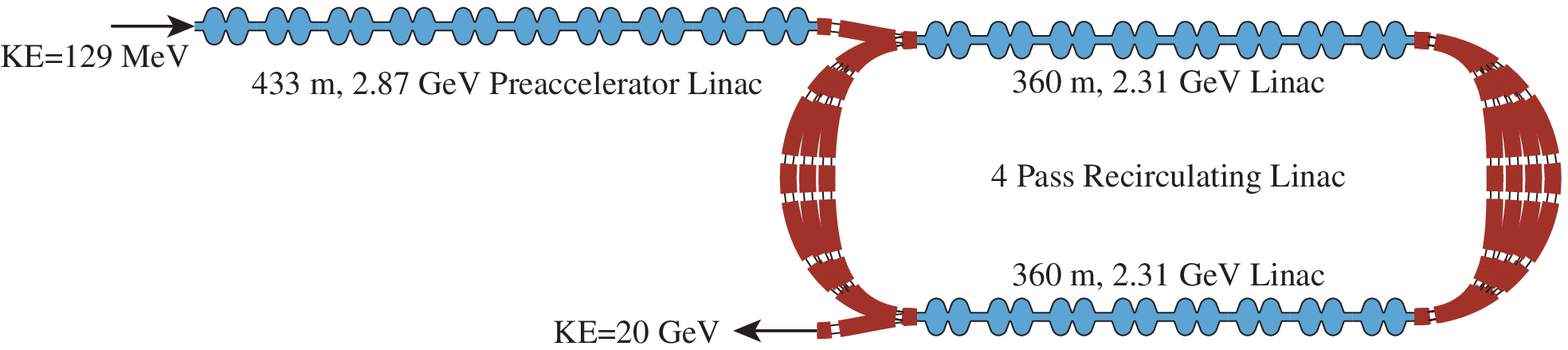}
  \caption{Accelerating system layout.}
  \label{fig:acc:layout}
\end{figure}

We intend to accelerate the muons to $\approx$ 2.87 GeV using a
preaccelerator linac and then on to 20 GeV using a recirculationg
linac (RLA) as shown in figure~\ref{fig:acc:layout}. The cost of this
stage is estimated to be \$220-250M for the pre-accelerator stage and
\$1250-1350M for the RLA. The acceleration stage is thus the most
costly segment of the Neutrino Factory and efforts are being made to
bring the cost down by examining other schemes such as rapid cycling
synchrotrons~\cite{summers} 
and also Fixed Field Alternating Gradient (FFAG) machines~\cite{johnstone}
that have large momentum acceptance and may need less cooling. 

Preliminary storage ring designs have been made with long straight 
sections that point towards detectors.

This completes the review of components used to build the Neutrino
Factory.  The amount of cooling needed for a Neutrino Factory is
significantly less than that needed to operate a Muon Collider. In
particular, a Muon Collider needs longitudinal as well as transverse
cooling. At the end of the linear cooling section, the transverse and
longitudinal emittances of the beams are 2.6 mm-rad and 30 mm-rad. For
the Muon Collider operating as a Higgs Factory, one needs to reduce
these to 0.14mm-rad and 9 mm-rad respectively. In addition, one needs
to be able to measure the energy of each muon bunch to a part in
$10^6$, so that one can scan the width of a narow Higgs resonance. It
has been demonstrated that using $g-2$ spin precession, it is possible
to measure the energy of the muon bunches to a precision comparable to
this provided the bunches retain a modest amount of polarization
($\approx$ 0.1)~\cite{rajatol}.  

In order to reduce the longitudinal
emittance, the collaboration has come up with an innovative
series of machines, called ``ring coolers'' which can be made to cool
in all 6 dimensions simultaneously.
\section{Ring Coolers}
\begin{figure}[tbh!]
\vspace{15mm}
\begin{minipage}[tbh!]{0.47\linewidth}
\includegraphics[width=\linewidth]{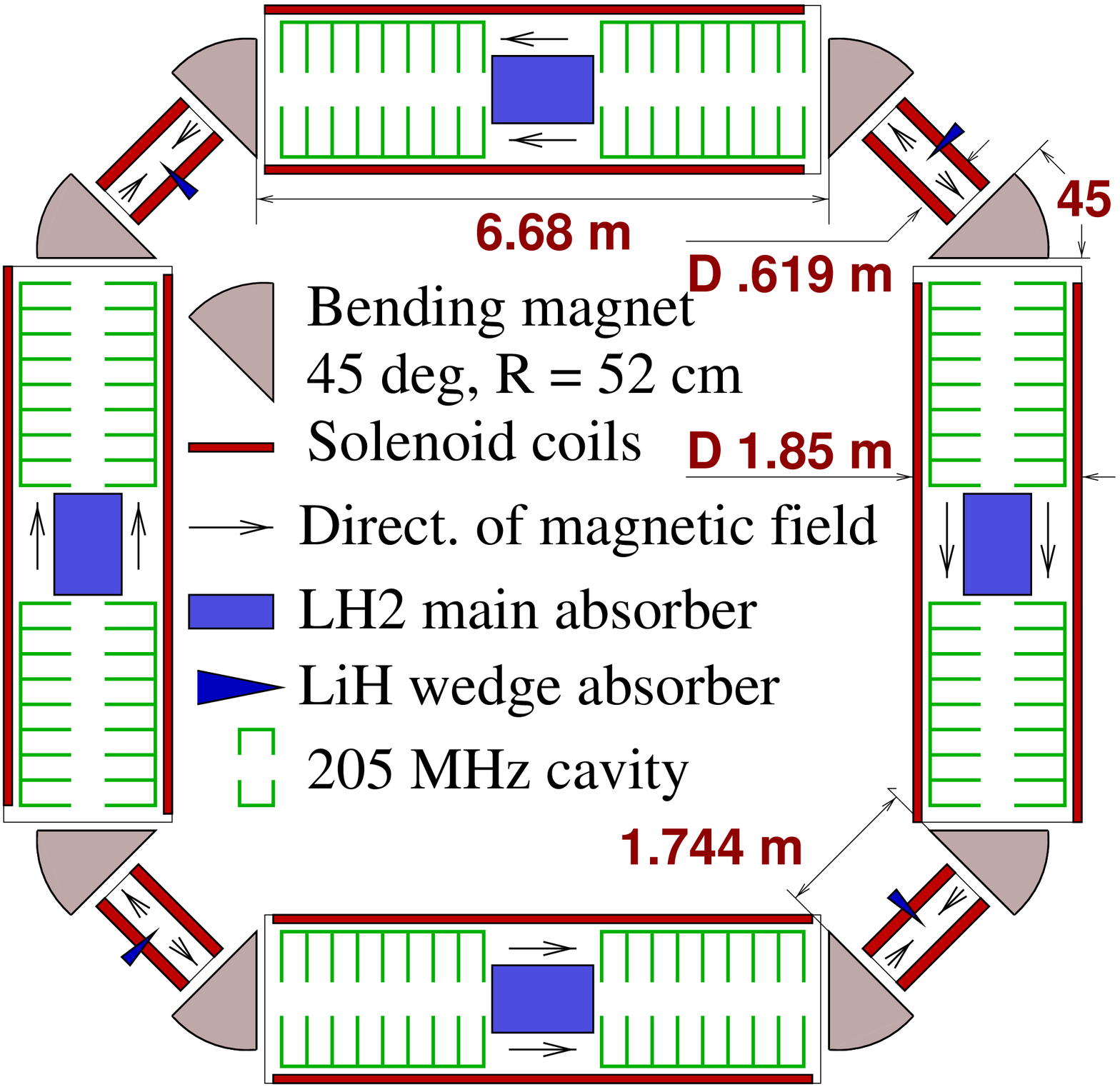}
\end{minipage}
\begin{minipage}[tbh!]{0.47\linewidth}
\begin{tabular}{ll}
Circumference & 36.963 m \\ Nominal energy at short & \\  straight section & 250 MeV
\\ Bending field & 1.453 T \\ Norm. field gradient & 0.5 \\
Max. solenoid field & 5.155 T \\ rf frequency & 205.69 MHz \\
Accelerating gradient & 15 MeV/m \\ Main absorber length & 128 cm \\
LiH wedge absorber & 14 cm \\ Grad. of energy loss & 0.75 MeV/cm \\
\end{tabular}
\end{minipage}
\caption{Layout and parameters of the solenoid based Tetra ring cooler
\label{ring}}
\end{figure}

The first such ring cooler scheme to be proposed, known as the Tetra ring, 
is shown~\cite{balbekov} in
Figure~\ref{ring}. In each of the four sides of the ring are placed
long solenoids whose fields reverse from side to side. The solenoidal
field increases to 5.15T at its center, where a liquid hydrogen
absorber is placed. The beam is bent in a circle by means of 8 special
dipoles (with field index $\frac{-1}{2}$) which bend as well as focus the beam
and cause dispersion in the middle of the short solenoids where the
lithium hydride wedge absorbers are placed. The wedge absorbers cause
a reduction in longitudinal emittance since the faster particles are
made to traverse through thicker portions of the wedge.

%
\begin{figure}[tbh!]
\begin{minipage}[t!]{0.47\linewidth}
\includegraphics[width=1.15\linewidth]{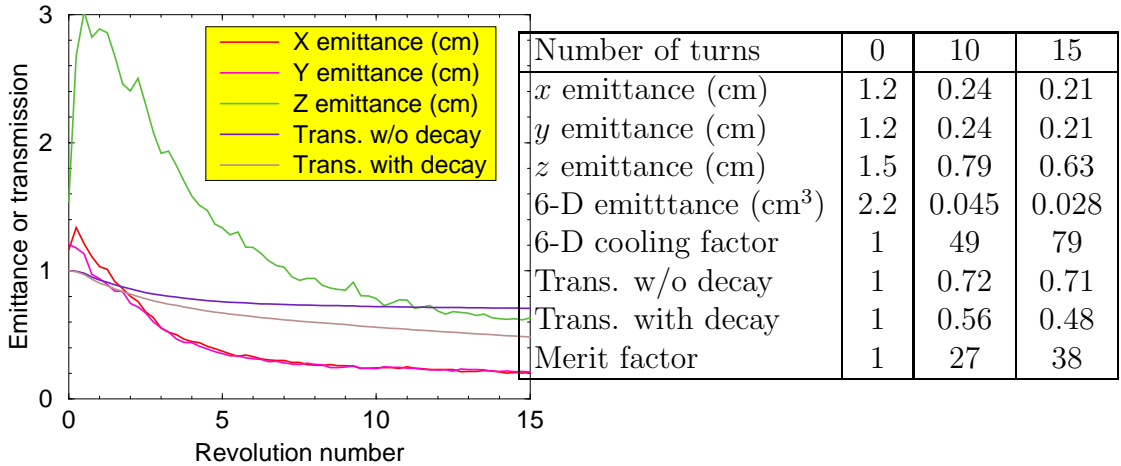}
\end{minipage}
\begin{minipage}[t!]{0.47\linewidth}
\vspace{-5mm}
\begin{tabular}{|l|c|c|c|}
\hline
Number of turns & 0 & 10 & 15 \\
\hline
$x$ emittance (cm) & 1.2 & 0.24 & 0.21 \\ $y$ emittance (cm) & 1.2 & 0.24
& 0.21 \\ $z$ emittance (cm) & 1.5 & 0.79 & 0.63 \\ 6-D emitttance (cm$^3$) &
2.2 & 0.045 & 0.028 \\ 6-D cooling factor & 1 & 49 & 79 \\ Trans. w/o
decay & 1 & 0.72 & 0.71 \\ Trans. with decay & 1 & 0.56 & 0.48 \\
Merit factor & 1 & 27 & 38 \\
\hline
\end{tabular}
\end{minipage}
\vspace{-3mm}
\caption{Evolution of the beam emittance/transmission at the TETRA ring cooler.
\label{evol}}
\vspace{-3mm}
\end{figure}
\begin{figure}[htb!] 
\includegraphics[width=\linewidth]{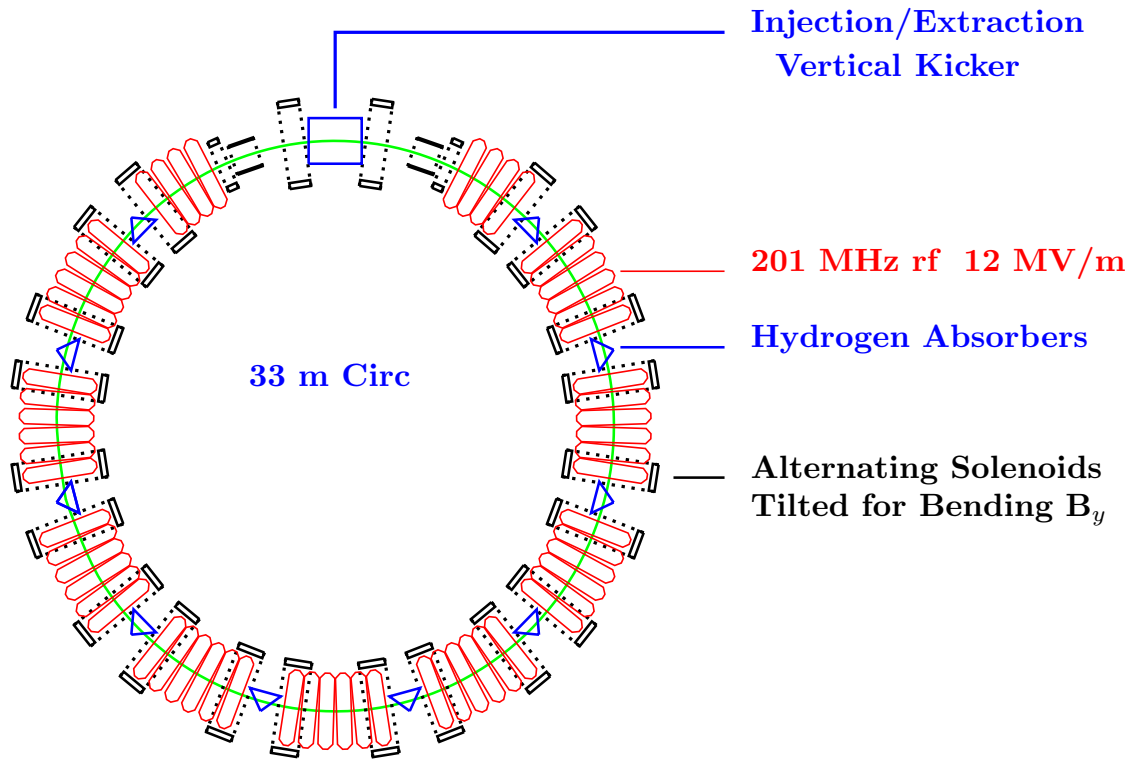} 
\caption{Layout of an RFOFO cooling ring. \label{rforing}}
\end{figure}
\begin{figure}[tbh!] 
\includegraphics[width=\linewidth]{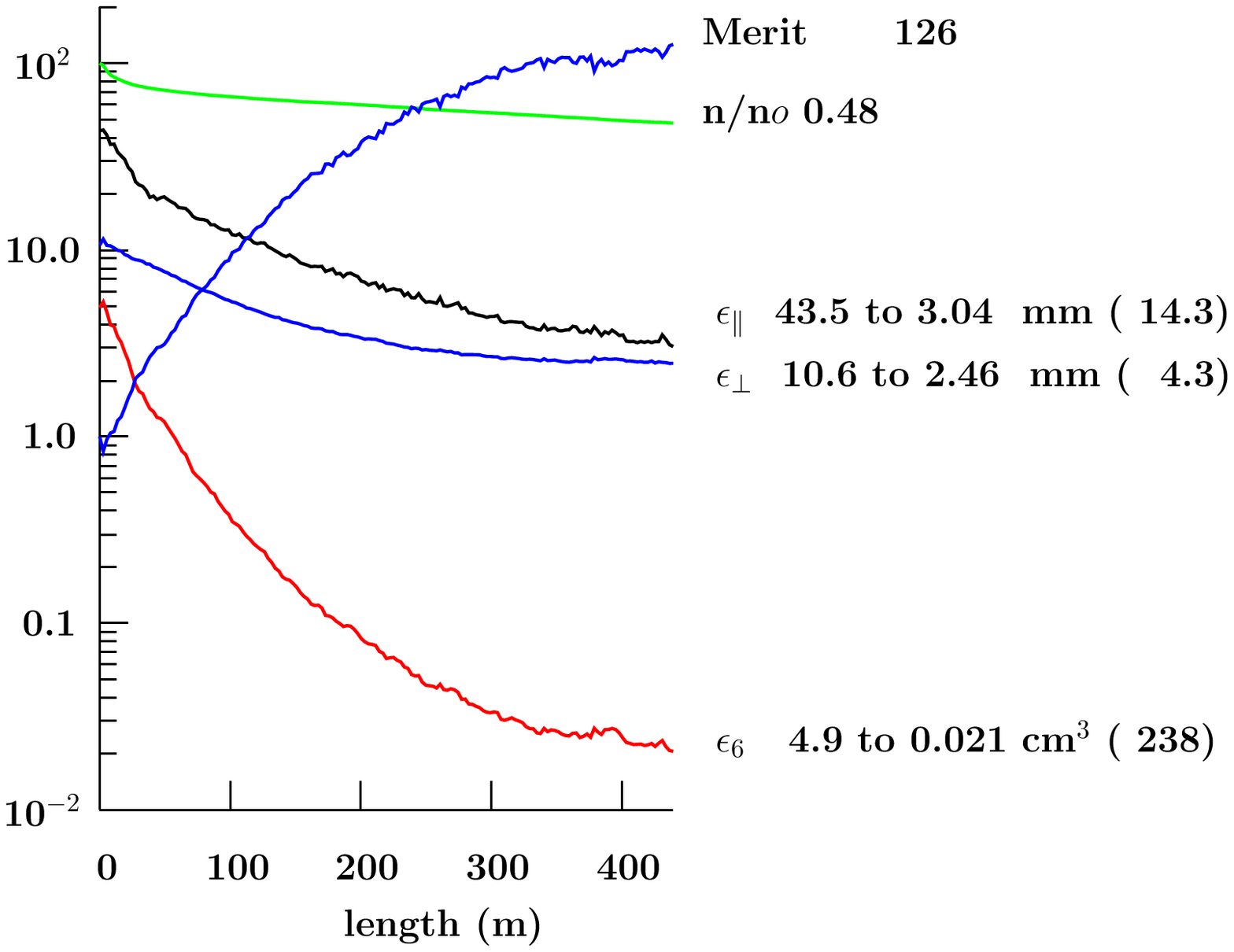} 
\caption{Transmission,
normalized transverse emittance, normalized longitudinal emittance,
normalized 6-dimensional emittance, and the merit factor, as a
function of distance in the RFOFO ring.  \label{all}}
\end{figure}
Figure~\ref{evol} shows the evolution of the transverse and
longitudinal emittance in this ring cooler as a function of turn
number. We were able to obtain 6D cooling in this ring simulation for
the first time. It must be pointed out however that the fields used
for the solenoids and dipoles were approximated by hard-edge
approximation. Calculations are under way to show that the cooling
persists when more realistic fields are used.

We have also simulated a version of the ring cooler~\cite{palmer},
where solenoids similar to those used in the linear channel are put in
a circle with wedge absorbers in between. Figure~\ref{rforing} shows
the layout of such a ring, known as the RFOFO ring, with a region  marked
out for injecting and extracting the beam.
Figure~\ref{all} shows the 6-D cooling and transmission as a function
of the circumference traversed in the ring. A measure of the efficacy
of the cooling is made by use of the so-called ``merit factor'', which
is defined as the ($\frac{6D\:emittance\:in\times transmission} {6D\:
emittance\: out}$) of the ring.
The major difficulty with these rings is the problem of injection and
extraction. In order to inject an uncooled beam, kickers need to be
fabricated that are orders of magnitude more powerful than any in
existence, since the kicker strength goes as the square of the transverse
emittance of the beam. The problem of injection of the beam is further
exacerbated by the need for long straight sections, that perturb the optics.

It was considerations of the problem of injection among others that
led to the  proposal~\cite{kirk} a third type of ring cooler, that uses
dipoles to bend and focus the beam. Recently, promising results have
emerged from simulations of such a cooler and discussions are underway
to see if such a cooler could be built cheaply to demonstrate the
concept of ring coolers.

The road to the Muon Collider will require a series of ring coolers, the 
last one will probably use lithium lenses to achieve 
very low emittances~\cite{fukui}.

\section{Conclusions}
Neutrino oscillations represent an exciting area of particle physics
beyond the standard model. The staged approach to Neutrino Factories
and the Muon Collider outlined here represent a means to provide a
diverse and rich ``base'' program of particle physics that will keep
any laboratory wise enough to invest in it at the forefront of physics
for years ahead.

\section{Acknowledgements}
The author wishes to thank the organizers of the NO-VE 2003 conference for
putting together an excellent collection of talks and discussion.

\end{document}